\documentclass[preprintnumbers,showpacs,amsmath,amssymb]{revtex4}
\usepackage{graphicx}
\usepackage{bm}
\begin{document}
\bibliographystyle{apsrev}

\title{Three types of spectra in one-dimensional systems with
random correlated binary potential}

\author{O. V. Usatenko}
\email{usatenko@ire.kharkov.ua}
\author{S. S. Melnik, V. A. Yampol'skii}

\affiliation{A. Ya. Usikov Institute for Radiophysics and Electronics \\
Ukrainian Academy of Science, 12 Proskura Street, 61085 Kharkov,
Ukraine}
\author{M. Johansson, L. Kroon, and R. Riklund}

\affiliation{Department of Physics and Measurement Technology, IFM
Link\"{o}ping University
 SE-581 83 Link\"{o}ping
 Sweden}

\begin{abstract}
The stationary one-dimensional tight-binding Schr\"{o}dinger
equation with a weak diagonal long-range correlated disorder in the
potential is studied. An algorithm for constructing the discrete
binary on-site potential exhibiting a hybrid spectrum with three
different spectral components (absolutely continuous, singular
continuous and point) ordered in any predefined manner in the region
of energy and/or wave number is presented. A new approach to
generating a binary sequence with the long-range memory based on a
concept of \emph{additive} Markov chains (Phys. Rev. E 68, 061107
(2003)) is used.
\end{abstract}


\maketitle
In recent years, the problem of transport of electro-magnetic
waves (and various excitations in solids) in one-dimensional (1D)
systems with the long-range correlated disorder has attracted much
attention~\cite{IK, T, IM}. The important significance of this
problem is due to exciting results that revise a commonly accepted
belief that any randomness (no matter how weak the randomness is)
in 1D structures results in the Anderson localization~\cite{An}.
It was also believed that 1D systems could not display a complex
dynamic feature such as the metal-insulator transition, which
gives rise to an appearance of the mobility edges. However, in
Refs.~\cite{Fl,Bo}, a highly nontrivial role of correlations was
shown, the divergency of the localization length for some specific
values of energy was observed. Using a perturbative approach, it
was also shown in Ref.~\cite{IK} that the position of the mobility
edges and the windows of transparency can be controlled by the
form of the binary correlator of a scattering potential, which was
supposed to be continuous, of a long-range and Gaussian. If the
correlations are of a long-range, i.e. they decay by the power
law, a continuum of extended states may appear in the energy
spectrum. In other words, the long-range correlations are
necessary for the observation of the metal-insulator transition
and co-existence of different type of spectra. A quite simple
method was used therein to construct a random potential with a
long-range correlation. From the experimental point of view, the
importance of recently obtained results may be explained by a
strong impact upon the creation of a new class of electron devices
and electromagnetic waveguides, which can be used as window
filters having new transmission properties.

In this paper we point out a new method for constructing the long
range correlated sequences of a two-valued site potential
$\varepsilon(n)$ with a given correlator and prescribed
probability distribution function (PDF), not only Gaussian. For
this potential the method given in~\cite{IK} does not work. An
attempt to construct a correlated dichotomous sequence where the
metal-insulator transition can be observed was made in
Ref.~\cite{CBIS}. Later, Ref.~\cite{CBIS2}, this result was
retracted.

We study a stationary one-dimensional tight-binding Schr\"{o}dinger
equation (the Anderson model~\cite{An}),
\begin{equation}
\label{1} \psi_{n+1}+\psi_{n-1}+(E-\varepsilon(n))\psi_{n}=0,
\end{equation}
with a site potential $\varepsilon(n)$ taking on two different
values $ \varepsilon_{0}$ and $ \varepsilon_{1}$.

Equation~(\ref{1}) is a prototype model describing a propagation
of excitations (electromagnetic waves, electrons, photons,
phonons) in deterministic ordered or random disordered systems
(solids or layered super-lattices). The type of ordering, i.e.,
the correlations in the site potential $\varepsilon(n)$ determine
the spectrum of excitation. The wave functions of excitations in
such systems are usually characterized by the Lyapunov exponent
$\Lambda$. Some typical examples of sequences having different
types of spectra are (spectrum is a set of allowed energies of
excitations in the system):

1. A system with a regular periodic variation of the site
potential $\varepsilon(n)$ possess an \emph{absolutely continuous}
(AC) spectrum and describes extended, delocalized states having
bounded wave functions with the Lyapunov exponent $\Lambda=0$.

2. A system with a random non-correlated potential
$\varepsilon(n)$ has the allowed states with the exponentially
localized wave functions and displays a \emph{point} (discrete)
spectrum with the positive Lyapunov exponent (the Anderson
localization).

3. A potential constructed using the deterministic quasi-periodic
Fibonacci chain (see the definition given below, Eq.~(\ref{6c}))
exhibits a \emph{singular continuous} (SC) Cantor-like spectrum
characterized by the power-law localized wave functions with
$\Lambda=0$. The singular continuous spectrum differs essentially
from the absolutely continuous one. This spectrum corresponds to a
singular continuous integrated density of states (the number of
states having the energies smaller then a given one). The
integrated density of states is the continuous function of energy,
but its derivation equals to zero almost everywhere. In terms of
the forbidden and allowed energy zones, this means that the
spectrum consists almost completely of forbidden gaps with the
infinite number of allowed energies between them. In other words,
the spectrum of quasi-periodic potential sequence possesses a
fractal structure.

In each of these structures only one pure type of spectrum is
presented. A nontrivial result of co-existence of continuous and
point spectra in one dimensional chain with correlated disorder
was presented in Ref.~\cite{IK}.

Correlated properties in the site potential $\varepsilon(n)$ are
determined by the pair correlation function,
\begin{equation}
\label{2} C(r)=<\varepsilon(n)\varepsilon(n+r)>_{s}-
<\varepsilon(n)>_{s}^{2},\qquad
<f(\varepsilon(n))>_{s}=\lim_{N\rightarrow \infty}\frac{1}{N}
\sum_{n=1}^{N}f(\varepsilon(n)),
\end{equation}
where the Cezaro average $<\cdot>_{s}=\lim_{N\rightarrow
\infty}\frac{1}{N} \sum_{n=1}^{N}\cdot$ can be treated as spatial
one. In the Born approximation the Lyapunov exponent is expressed,
see Ref.~\cite{Ta,GF,L,IK}, in terms of the Fourier transform
$\widetilde{K}$ of this two-point correlation function,
\begin{equation}
\label{3}
\Lambda(E)=\frac{(\varepsilon_1-\varepsilon_0)^2\widetilde{K}(2k)}{32\,\sin^{2}k}
, \qquad k\in [-\pi, \pi] , \qquad K(r)=\frac{C(r)}{C(0)}.
\end{equation}
The correlation function $K(r)$ and its Fourier transform are
connected by the following relations,
\begin{equation}\label{3a}
\widetilde{K}(k)=1+2\sum_{r=1}^{\infty}K(r)\cos(kr), \, \,
K(r)=\frac{1}{\pi}\int_{0}^{\pi}\widetilde{K}(k)\cos(kr)dk
\end{equation}
Here the evenness of the functions $K(r)$ and $\widetilde{K}(k)$
is taking into account. The energy $E$ of the eigenstate and the
wave number $k$ in Eq.~(\ref{3}) in zero approximation on the
strength of disorder are related by the simple formula,
\begin{equation}
\label{3a} E=-2\cos k.
\end{equation}
Thus, there exists the relation between the correlations in the
on-site potential $K(r)$ and localization properties of
eigenstates expressed in terms of the Lyapunov exponent
$\Lambda(E)$ (or by means of integrated density of states). This
observation enables one to construct random correlated sequences
with prescribed spectral properties and provides a recipe for
designing filters of arbitrary complexity. Starting from a
desirable spectral dependence $\Lambda(E)$ (or integrated density
of states) we have to solve the inverse problem of construction a
sequence of "symbols" $\varepsilon(n)$. This program was partially
implemented in Ref.~\cite{IK}, where the one-dimensional
tight-binding Schr\"{o}dinger equation with \emph{two} kind of
spectra (absolutely continuous and point) and Gauss distribution
of the site potential was examined. However, it is
known~\cite{RS}, that in the general case the space of states can
be decomposed into the direct sum of subspaces with the wave
functions belonging to the point, singular continuous and
absolutely continuous parts of spectra. This statement is closely
connected with a possibility for presenting an integrated density
of states, as for any monotone function, in the sum of three
functions: the stepwise, singular and absolutely continuous ones.
Here we point to a general method, which is the extension of that
used in Ref.~\cite{IK}, and construct a particular example of 1D
system having all three above-mentioned types of
spectra~\cite{Note}.

To this end we use a new instrument of constructing the correlated
sequence of $\varepsilon(n)$ with a given correlation function
$K(r)$ using an \emph{additive} Markov chain~\cite{uya,uyakm}. Here
we give a short description of this method.

Consider a homogeneous binary sequence of symbols,
$\varepsilon(i)=\{\varepsilon_{0},\varepsilon_{1}\}$, $i\in
\textbf{Z} =...,-2,-1,0,1,2,...$. To determine the
$N$-\textit{step Markov chain} we have to introduce the
\emph{conditional probability} function $P(\varepsilon(i)\mid
\varepsilon(i-N),\varepsilon(i-N+1),\dots ,\varepsilon(i-1))$. It
is a probability of occurring the definite symbol $\varepsilon_i$
(for example, $\varepsilon(i) =\varepsilon_{1}$) after $N$-word
$T_{N,i}$, where $T_{N,i}$ stands for the sequence of symbols
$\varepsilon(i-N),\varepsilon(i-N+10),\dots ,\varepsilon(i-1)$.
The \emph{additive} Markov chain is characterized  by the
conditional probability function of the form
\begin{equation}
\label{4} P(\varepsilon(i)=\varepsilon_{1}\mid
T_{N,i})=p_1+\sum\limits_{r=1}^{N}
F(r)(\varepsilon(i-r)-<\varepsilon>).
\end{equation}
We refer to the amount $F(r)$ (introduced first in
Ref.~\cite{mel}) as the \emph{memory function}. It describes the
strength of influence of previous symbol $\varepsilon(i-r)$
$(r=1,...,N)$ upon a generated one, $\varepsilon(i)$. Here $p_1$
is the relative part of symbols $\varepsilon_{1}$ among  the total
number of symbols in the whole sequence ($p_1$ is not necessarily
equal to $1/2$). There is a single-valued relation between the
memory function $F(r)$ and the correlation function $K(r)$ of the
Markov chain, see Ref.~\cite{mel,melg},
\begin{equation}
\label{5} K(r)=(\varepsilon_{1}-\varepsilon_{0})
\sum\limits_{r'=1}^{N}F(r')K(r-r'), \quad r \geq 1.
\end{equation}
This equation allows finding the memory function $F(r)$ and
effective constructing the Markov chain with the obtained
conditional probability $P(.\mid .)$ introduced in Eq.~(\ref{4}).

Let us present a ''\emph{Recipe}'' and enumerate the consecutive
steps of the general method for constructing the correlated
sequence of $\varepsilon_{n}$ with desired spectra:
\begin{enumerate}
    \item Dividing the interval of energy $E$ (or, that is
    the same, of the wave number
    $k$) to the regions where we want to have different types of spectra.
    \item Prescribing $\Lambda$ or/and $\widetilde{K}(2k)$ in these
intervals, we construct ''by hand'' the total Fourier transform of
correlation function $\widetilde{K}(2k)$ for all values of $k\in
[-\pi, \pi]$.
    \item Calculating $K(r)$ as inverse transform of
    $\widetilde{K}(2k)$.
    \item Solving Eq.~(\ref{5}), we determine the  memory function $F(r)$
    and the conditional probability function $P(.\mid .)$.
    \item Constructing sequence of $\varepsilon_{0}$ and
$\varepsilon_{1}$ with obtained function $P(.\mid .)$.
\end{enumerate}

\begin{figure}[ht!]
{\includegraphics[width=0.40\textwidth,height=0.30\textwidth]
{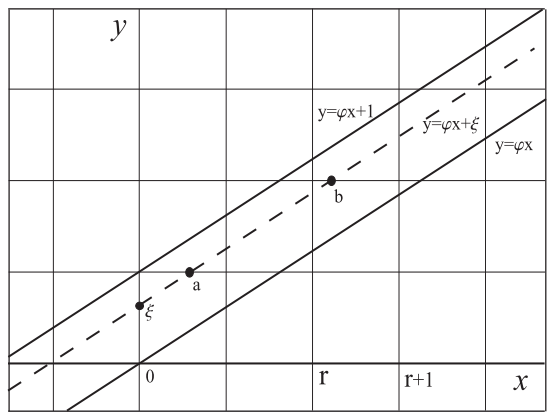}} \caption{The line  $y=\varphi x + \xi $ ''generates''
the Fibonacci sequence~(\ref{6a}). The strip bounded by two lines
$y=\varphi x $ and $y=\varphi x + 1$ corresponds to an ensemble of
all possible Fibonacci sequences with $0\leq \xi < 1$. The line
$y=\varphi x + \xi $ intersects a horizontal line $y=1$ between 0
and 1 (and so does another line $y$ between $r$ and $r+1$) at point
a (and $b$) and gives $\Phi_{\xi}(0)=1$ (and $\Phi_{\xi}(r)=1$).
Such values of $\xi $ give a nonzero contribution to the
integral~(\ref{7}).} \label{f1}
\end{figure}

To construct the total correlation function $K(r)$ we have to
calculate the "patrial" ones. Let us present here some simple
examples of the correlation functions corresponding to the
above-mentioned sequences exhibiting the pure spectra of periodic,
random and quasi-periodic chains.

It follows from definition~(\ref{2}) that the correlation function
of a chain of alternating potentials
$\varepsilon(n)=...\varepsilon_{0}\varepsilon_{1}\varepsilon_{0}
\varepsilon_{1}...$ has the form:
\begin{equation}\label{6}
C_{a,2}(r)=\frac{(-1)^r}{4}(\varepsilon_{1}-\varepsilon_{0})^2.
\end{equation}

It follows from the same definition~(\ref{2}) that the correlation
function of the sequence with the same potentials
$\varepsilon(n)=...\varepsilon_{0}\varepsilon_{0}
\varepsilon_{0}...$ is equal to zero for all distances $r$.
Nevertheless, the ratio $C(r)/C(0)=K(r)$ should be defined as
unity. Only in this case Eq.~(\ref{3}) gives the correct value of
the Lyapunov exponent.

The correlation function of the non-biased non-correlated random
sequence of $\varepsilon_{n}$ is
\begin{equation}\label{6a}
C_p(r)=(\varepsilon_{1}-\varepsilon_{0})^2 \left\{%
\begin{array}{ll}
   1/4, \ \ r = 0 \\
   0, \ \ r \neq 0. \\
\end{array}%
\right.
\end{equation}
Let us derive the correlation function of the quasi-periodic chain
of potentials $\varepsilon_{n}$ constructed using the Fibonacci
chain,
\begin{equation}\label{6c}
\Phi_{\xi}(r)=[(r+1)\varphi+\xi]-[r\varphi+\xi].
\end{equation}
as,
\begin{equation}\label{6b}
\varepsilon_{r}(\xi)=\varepsilon_{0}+(\varepsilon_{1}-\varepsilon_{0})
\Phi_{\xi}(r), \qquad \qquad \xi\in[0,1),
\end{equation}
\begin{figure}[ht!]
{\includegraphics[width=0.45\textwidth,height=0.36\textwidth]
{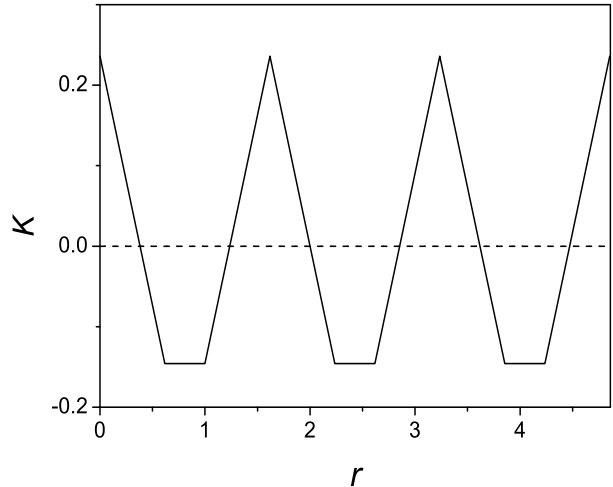}} \caption{The function $K_s(r)=C_s(r)/C_s(0)$
determined by Eq.~(\ref{8}) vs. the \emph{continuous} variable $r$
for $\varepsilon_0 = 0$, $\varepsilon_1 = 1$.} \label{f2}
\end{figure}
Here $\varphi$ is the so-called golden number, $\varphi =(\sqrt5
-1)/2$, and $\xi$ selects some concrete sequence from among all
possible Fibonacci sequences. The geometrical meaning of this
construction is explained in Fig.~\ref{f1}. To determine the
correlation function of this sequence it is convenient to use an
ergodic theorem~\cite{AA} about the homogeneity of distribution of
$y_{n}=n\varphi - [n\varphi]$ on the interval $0\leq y \leq 1$ for
any irrational number $\varphi$ and calculate correlation function
$C(r)$ using ensemble average $<\cdot>_{a}$ instead of space
average~(\ref{2}),
\begin{equation}
\label{7} C(r)=<\varepsilon_{n}\varepsilon_{n+r}>_{a}-
<\varepsilon_{n}>_{a}^{2}\qquad <f(\varepsilon_{n})>_{a}=
\int_{0}^{1}f(\varepsilon_{n}(\xi))d\xi.
\end{equation}

From a simple geometric consideration, see explanation in
Fig.~\ref{f1}, we obtain the following result,
\begin{equation}
\label{8} C_s(r) = (\varepsilon_{1}-\varepsilon_{0})^2 \left \{
\begin{array}{ll}
2\varphi - 1 - \{\varphi r \}, & \{\varphi r\} \leq 1 - \varphi, \\
3\varphi - 2, & 1 - \varphi \leq \{\varphi r\} < \varphi, \\
2\varphi - 2 + \{\varphi r \}, & \varphi \leq \{\varphi r\}.
\end{array} \right.
\end{equation}

\begin{figure}[h!!!]
{\includegraphics[width=0.45\textwidth,height=0.36\textwidth]
{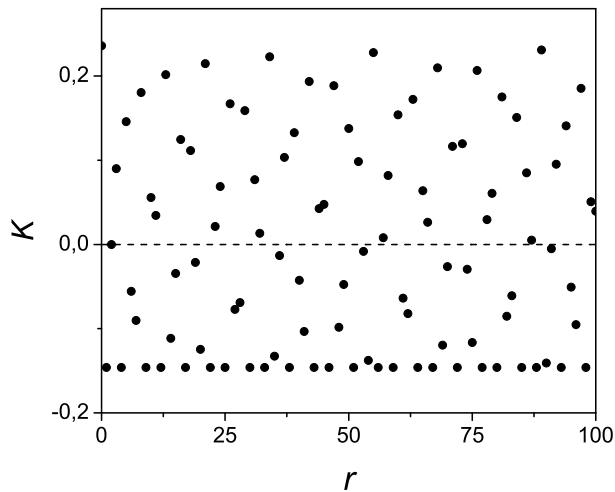}} \caption{The correlation function
$K_s(r)=C_s(r)/C_s(0)$ of the Fibonacci chain~(\ref{6b}) for
$\varepsilon_0 = 0$, $\varepsilon_1 = 1$. } \label{f3}
\end{figure}

\begin{figure}[ht!]
{\includegraphics[width=0.40\textwidth,height=0.30\textwidth]{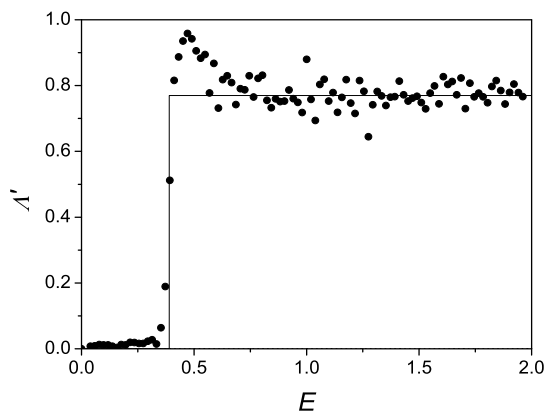}}
\caption{The ''normalized'' Lyapunov exponent $\Lambda'$ of the
binary correlated system vs. the energy $E$. The solid line is the
prescribed characteristic of system with the stepwise Lyapunov
exponent $\Lambda'$. The dots are the result from the calculation of
the Lyapunov exponent of the sequence $\varepsilon(n)$ constructed
by means of the memory function $F(r)$ obtained by numerical
solution of Eq.~(\ref{5}) for the correlation function
$K(r)=\sin(\pi r/3)/(\pi r/3)$, corresponding to the stepwise
Lyapunov exponent.} \label{f4}
\end{figure}

Here $\{x\}$ is the fractional part of $x$. The function
$K_s(r)=C_s(r)/C_s(0)$ of continuous argument $r$ is shown in
Fig.~\ref{f2}. Actually, the correlation function is defined for
the integer arguments, which values are incommensurable with the
period $1/ \varphi$ of function $C_{s}(r)$. The function
$C_{s}(r)$ of the discrete argument is shown in Fig.~\ref{f3}.
Numerical calculations of $C_{s}(r)$ carried out using spatial
averaging~(\ref{2}) exhibit a good agreement with the analytical
one~(\ref{8}).
Below we construct an example of the sequence with a prescribed
spectral property. Let us suppose that the Schr\"{o}dinger
Eq.~(\ref{1}) has all three kind of spectra. Let $k_{i1}$ and
$k_{i2}$ be the wave vectors corresponding to the top and bottom
edges of the bands where the singularly continuous, $i=s$,
absolutely continuous, $i=a$, and point (discrete), $i=p$, spectra
take place. In fact, we do not need to calculate Fourier
transforms $\widetilde{K}(k)$ of the correlation function as it is
indicated above in the paragraph 3 of the ''\emph{Recipe}''. We
can calculate the Fourier transform of the composed correlation
function $K(r)$ presented in $k$-space via characteristic
functions $\chi_{[k_{i1},k_{i2}]}(k)$ of the intervals
$[k_{i1},k_{i2}]$,
\begin{equation}
\label{8a}
\widetilde{K}(k)=\sum_{i=\{a,s,p\}}\chi_{[k_{i1},k_{i2}]}(k)
\widetilde{K}_{i}(k),
\end{equation}
where characteristic function $\chi_{[k_{1},k_{2}]}(k)$ is equal
to $1$ if $k\in [k_{1},k_{2}]$ and equals $0$ otherwise.
Straightforward calculations yield:
\begin{equation}
\label{9} K(r)=\sum_{i}\left[ \frac{k_{i2}-k_{i1}}{\pi}K_{i}(r)+
\frac{1}{\pi} \sum_{r'=1}^{\infty}[K_{i}(r+r')+
K_{i}(r-r')]\frac{\sin r' k_{i2}-\sin r' k_{i1}}{r'} \right].
\end{equation}
Here the functions $(\sin r' k_{i2}-\sin r' k_{i1})/r'$ are the
Fourier transforms of the characteristic functions. In the case
$i=p$, the contribution of the point spectrum to Eq.~(\ref{9}) is
reduced to one term only,
\begin{equation}
\label{10} [...] =\left \{
\begin{array}{ll}
\dfrac{\sin r k_{p2}-\sin r k_{p1}}{\pi r}, & r\neq 0, \\
\dfrac{k_{i2}-k_{i1}}{\pi}, & r=0.
\end{array} \right.
\end{equation}
Note that in deriving Eq.~(\ref{9}) we did not use the concrete form
of correlation functions Eqs.~(\ref{6}),~(\ref{6a}) and~(\ref{8}).
In Fig.~\ref{f4} the result of numerical simulation of the
normalized Lyapunov exponent $\Lambda'(E)=32\, \sin^2(k)\
\Lambda(E)/(\varepsilon_1-\varepsilon_0)^2$ is presented for the
system exhibiting two types of spectrum, namely, absolutely
continuous with $\Lambda'=0$ at $0<E<1$ and point spectrum
$\Lambda'=1.5$ at $1<E<2$~\cite{Note2}.

Thus, we have proposed a method for constructing 1D sequences of
sites $\varepsilon(n)$ with a correlated disorder exhibiting a
hybrid spectrum with three different spectral components ordered in
any predefined manner in the region of energy $E$ and/or wave number
$k$.

Authors are grateful to L. A. Pastur for the discussion which was
helpful in the formulation of the problem under study. We thank S.
A. Gredeskul, A. A. Krokhin, and F. M. Izrailev for useful and
enlightening discussions.


\begin{thebibliography}{20}
\bibitem{Note} In the frame of the theory of perturbation
one cannot pretend that in each domain of energy the spectrum
contains only one pure component. We can claim only that in each
energetic region there dominate one of spectral component.
\end{thebibliography}
\end{document}